\documentclass[a4paper]{article}

\usepackage{arxiv}
\usepackage{todonotes}

\usepackage[utf8]{inputenc} 
\usepackage[T1]{fontenc}    
\usepackage{hyperref}       
\usepackage{url}            
\usepackage{booktabs}       
\usepackage{amsfonts}       
\usepackage{nicefrac}       
\usepackage{microtype}      
\usepackage{todonotes}

\newcommand{\calA}{\mathcal{A}}
\newcommand{\calU}{\mathcal{U}}
\newcommand{\calC}{\mathcal{C}}
\newcommand{\AP}{\mathrm{AP}}
\newcommand{\pow}[1]{\mathcal{P}(#1)}
\newcommand{\Acc}{\mathrm{Acc}}

\title{The Extended HOA Format for Synthesis}

\author{
  Guillermo A. P\'erez\\
  Department of Computer Science\\
  University of Antwerp\\
  Middelheimlaan 1, Antwerpen 2020, Belgium\\
  \texttt{guillermoalberto.perez@uantwerpen.be}
}

\begin{document}
\maketitle

\begin{abstract}
    We propose a small extension to the Hanoi Omega-Automata
    format to define reactive-synthesis problems. Namely, we
    add a ``controllable-AP'' header item specifying
    the subset of atomic propositions which is
    controllable. We describe the
    semantics of the new format and propose an output format
    for synthesized strategies. Finally, we also comment on
    tool support meant to encourage fast adoption of
    the extended
    Hanoi Omega-Automata format for synthesis.
\end{abstract}


\section{Introduction}
The field of formal verification has studied several forms of model
checking of reactive systems~\cite{bk08,hmc18}. Model checking essentially
asks whether a model of a system satisfies a specification given
in some formal logic, e.g. linear temporal logic (LTL). Reactive
synthesis, or the automatic synthesis of a reactive system from its
specification, goes a step further by completely circumventing the
design of a system and its modelization. The earliest version of the
synthesis problem was stated by Alonzo Church~\cite{church62}. Since then,
several versions of it arising from different specification
formalisms and implementation restrictions have been studied.

The reactive synthesis competition
(SYNTCOMP,\footnote{See \url{http://www.syntcomp.org/}}
for short) was started in 2014
by R. Bloem and S. Jacobs to collect benchmarks and provide an objective
comparison of the state-of-the-art synthesis tools. Initially, the 
competition consisted of a realizability and a synthesis track for
safety specifications. The realizability track just asks whether a 
system satisfying a given specification exists while the synthesis
track asks for it to be output by the tool if it exists.

\paragraph{Synthesis from LTL.}
Since 2016 tracks with LTL (given in the TLSF~\cite{tlsf16} format)
specifications were introduced. Currently, the synthesis problem for such
specifications is usually done in two steps: First, the LTL specification
is translated into a deterministic infinite-word automaton. Second, a
game is played on the automaton between the controller and its environment.
A strategy for the controller in such a game corresponds to an implementation
which satisfies the LTL specification.

\paragraph{Synthesis from automata.}
While the Hanoi Omega-Automata (HOA) format~\cite{hoaf15} has arguably
helped LTL-to-automata translation
tools become objectively comparable, the same has
not yet occurred for game-solving tools aimed at solving LTL synthesis.
Furthermore, when comparing LTL-synthesis tools, it is not clear at which
one of the aforementioned steps they excel.
This is why in this document we propose an extension of the HOA format
which allows to directly encode games played on automata.

\section{Input format}
The Hanoi Omega-Automata (HOA) format is a format to describe finite-state
automata that accept (as their language) sets of infinite words. Such an
automaton $\calA$ is a tuple $(Q,q_0,\pow{\AP},\Delta,\Acc)$ where
\begin{itemize}
    \item $Q$ is a finite set of states,
    \item $q_0 \in Q$ is an initial state,
    \item $\AP$ is a finite set of atomic propositions and the alphabet $\pow{\AP}$ of the automaton consists of their valuations,
    \item $\Delta \subseteq Q \times \pow{\AP} \times Q$ is the transition relation, and
    \item $\Acc \subseteq (Q \cdot \pow{\AP})^\omega$ is an acceptance condition, i.e. a set of infinite runs of the automaton which are considered accepting.
\end{itemize}
The language of the automaton is defined as usual: the set of words for
which the automaton has an accepting run.

\subsection{Original format}
The original HOA format was described in a CAV'15 paper~\cite{hoaf15}. More
information, as well as Java and C++ parsers for the format,
can also be found in the website
\url{http://adl.github.io/hoaf/}.

Every HOA-format file is split into a \emph{header} and the \emph{body}
of the automaton. The body encodes all the transitions $\Delta$
of the automaton $\calA$. The header gives meta-information regarding,
amongst others, how $\Acc$ is given. Importantly, it also holds information
about the atomic-proposition set $\AP$.

\paragraph{Atomic propositions.}
One of the \emph{header items} is \texttt{AP}. This item gives
the number of atomic propositions followed by a space-separated list
of unique names for each atomic proposition. These are double-quoted
valid C-strings and implicitly numbered, starting from $0$, from left
to right.

\paragraph{Acceptance condition.}
In the HOA format, information about the acceptance condition is given in two parts. First,
an \emph{acceptance-condition name} is given
in the form of a string. Second, a Boolean
combination of conditions over \emph{acceptance
sets} gives the actual acceptance condition. 
Each transition can belong to several acceptance
sets.\footnote{For clarity, we are focusing on edge-labelled
automata. However, the HOA and extended HOA formats support
state-labelled automata as well.} Much like in
Muller automata~\cite{muller63}, 
the acceptance condition
then specifies which
runs are considered accepting by indicating which 
combinations of finitely and infinitely appearing
acceptance sets are ``good''. For instance, 
\begin{verse}
``transitions
from the first acceptance set can only appear finitely often
and transitions from the second acceptance set must appear
infinitely often.''
\end{verse}

\subsection{Extension for synthesis}
In the context of reactive synthesis it is necessary to make a distinction
between controllable and uncontrollable atomic propositions. That is, the
propositions whose value is set by the uncontrollable environment and
those whose value is set by the controller.
We propose a very simple extension of the HOA format, based
on the synthesis extension of AIGER~\cite{eaiger14}, to make this
distinction explicitly. We shall add a header item with the following specification

\begin{center}
    \texttt{header-item ::= "controllable-AP:" INT* | ...}
\end{center}

\noindent
In other words, the new header item will have a list of space-separated integer indices.
These indicate which atomic propositions from the 
\texttt{AP} header item  are controllable.
All other atomic propositions are implicitly assumed to be uncontrollable.

\section{Semantics: Games played on automata}
We now present the formal semantics of the synthesis problem when
the input specification is given in the form of an automaton.
For convenience, we assume that we are given a deterministic
automaton $\calA = (Q,q_0,\pow{\AP},\delta)$. Hence, $\delta$
can be thought of as a function $\delta : Q \times \calU \times
\calC \to Q$, for $\calU$ and $\calC$ valuations of the
uncontrollable and controllable sets of atomic propositions
respectively.

\paragraph{A game.}
The game we describe is played by a controller and its environment
for an infinite number of rounds.
In every round, the game starts from a state $q \in Q$.
The environment chooses a valuation $u \in \calU$, followed
by a choice of valuation $c \in \calC$ made by the controller.
The round then ends with an update of the current state from $q$
to $\delta(q,u,c)$. The initial state is $q_0$ and the \emph{play}
is winning for the controller if and only if an accepting run of
$\calA$ was constructed.

Formally, a \emph{strategy of the controller} is a function
$\sigma : (Q \cdot \pow{\AP})^* \cdot Q \times \calU \to \calC$
mapping the observed sequence of states and atomic-proposition
valuations to a valuation of the controllable propositions.
The strategy is \emph{winning} if and only if all plays consistent
with it yield an accepting run of $\calA$.

\paragraph{Realizability and synthesis.}
Given an automaton as input, the realizability problem asks whether
there exists a winning strategy for the controller. The synthesis
problem further asks for the winning strategy itself to be
output if it exists.

For a more detailed description of the reactive synthesis problem
and the game-theoretical approach to solving it we refer the
reader to~\cite{ag11,hmc18}.

\subsection{A note on non-determinism}
Note that
giving one of the players the power to resolve the non-determinism
does not, in general, yield a correct algorithm~\cite{gfg06}.
Indeed, when the input automaton $\calA$ is non-deterministic
then the
game has to be formalized in a more general way: The game played
between the controller and its environment is actually a
Gale-and-Stewart game~\cite{galestewart53} whose winning
condition for the controller
is the language of $\calA$. The game solver must either determinize
the given automaton or implement some algorithm that deals
with the non-determinism explicitly.

\section{Output format}
A strategy of the controller can be encoded in the original
AIGER format~\cite{aiger11}. \emph{All} the uncontrollable
atomic propositions should become inputs of the
encoded sequential circuit and \emph{all}
the controllable inputs should become outputs. \emph{Latches} can
be used to encode further information such as the states
of the input automaton.

\subsection{Why two formats?}
We could have chosen to encode strategies as automata in the HOA format, just
like the input automata. Nevertheless, in the spirit of this track
representing an intermediate step in the LTL-synthesis pipeline, we believe
having an output format which matches that of the LTL tracks is beneficial for
the competition. It is also easier for people to adopt tools participating in
this new track and integrate them with an LTL-to-automaton tool to obtain a
toolchain comparable to tools pariticipating in the LTL-synthesis tracks.

\section{Restrictions for the first edition}
In the context of SYNTCOMP 2020, we will restrict
the benchmark set to \textbf{deterministic and complete parity
automata} only.

For parity automata in the HOA format, the acceptance
condition is given
via two parameters: $m \in \{\max,\min\}$ and
$p \in \{0,1\}$. Additionally,
the transitions of the automaton are assumed
to be labelled by natural
numbers. For every run, we then consider the set
$\mathrm{Inf}$ of transition labels
appearing infinitely often. According
to the parity acceptance condition, the run is accepting if and only if
\[
    m(\mathrm{Inf}) \equiv p \pmod{2},
\]
e.g. the maximal label appearing infinitely often is even.

\section{What about the PGSolver format?}
The PGSolver collection of solvers and benchmarks provides a useful
framework to test and compare parity-game algorithms. To
streamline the adoption of the extended HOA format for synthesis,
we provide a simple transformation from it to the
PGSolver format. 

The \texttt{hoa2pg} tool\footnote{Available
from \url{https://github.com/gaperez64/hoa-tools}} is a simplified
version of one of the components present in a prototype
synthesis tool which translates linear temporal logic into
a parallel composition of several parity games~\cite{rp19}.
In essence, it translate an extended-HOA automaton
into a parity game using binary decision diagrams to
abstract the atomic-proposition valuations labelling the
transitions of the original automaton and by adding
intermediate vertices representing the choices of the
environment. \texttt{hoa2pg} can be directly used to
obtain a realizability solver together with a PGSolver-format
parity-game solver. 

\paragraph{Synthesis.}
As the transformation abstracts away
the atomic propositions, synthesis of strategies in the
AIGER format is less straightforward and may require
modifying \texttt{hoa2pg} to output more information about
the translation into the PGSolver format.

\section*{Acknowledgements}
We would like to thank Ayrat Khalimov, Salomon Sickert, and Swen Jacobs for
their valuable feedback on earlier versions of this document.

\bibliographystyle{unsrt}  
\bibliography{references}

\end{document}